# Toward Automated Detection of Biased Social Signals from the Content of Clinical Conversations

Feng Chen, MS[1], Manas Satish Bedmutha, MS[2], Ray-Yuan Chung, MPH[1], Janice Sabin, PhD, MSW[1], Wanda Pratt, PhD, FACMI[1], Brian R. Wood, MD[1], Nadir Weibel, PhD[2], Andrea L. Hartzler, PhD[1], Trevor Cohen, MBChB, PhD, FACMI[1]
[1]University of Washington, Seattle, WA; [2]University of California San Diego, La Jolla, CA

**Abstract**

*Implicit bias can impede patient-provider interactions and lead to inequities in care. Raising awareness is key to reducing such bias, but its manifestations in the social dynamics of patient-provider communication are difficult to detect. In this study, we used automated speech recognition (ASR) and natural language processing (NLP) to identify social signals in patient-provider interactions. We built an automated pipeline to predict social signals from audio recordings of 782 primary care visits that achieved 90.1% average accuracy across codes, and exhibited fairness in its predictions for white and non-white patients. Applying this pipeline, we identified statistically significant differences in provider communication behavior toward white versus non-white patients. In particular, providers expressed more patient-centered behaviors towards white patients including more warmth, engagement, and attentiveness. Our study underscores the potential of automated tools in identifying subtle communication signals that may be linked with bias and impact healthcare quality and equity.*

**Introduction**

The quality and social dynamics of patient-provider interactions in clinical encounters are significant factors in shaping patient outcomes. Notably, these interactions are negatively associated with provider trust among underrepresented racial and ethnic groups, which can affect patient satisfaction.[1] Patient-centered communication, characterized by responsiveness to patient health concerns, beliefs, and contextual variables, is associated with better health outcomes.[2] However, implicit bias may impede these interactions, leading to inequitable healthcare and outcomes. Implicit biases are "attitudes or stereotypes that affect our understanding, decision-making, and behavior, without our even realizing it."[3] They are particularly detrimental to patients from historically marginalized groups, leading to disparities in care quality and health outcomes.[4] Some healthcare providers have implicit race bias indicating more positive attitudes toward white patients than people of color.[4, 5] Such bias is often expressed through verbal and nonverbal patient-provider communication, such as speech and body language.[6] For example, Cooper and colleagues[7] found that clinicians with stronger implicit attitudes favoring White people over Black people, express greater verbal dominance (i.e., greater conversational control reflected by the ratio of provider to patient statements) and less patient-centeredness with Black patients compared with White patients.

Traditional methods of assessing patient-provider communication have predominantly relied on observational techniques and manual coding systems, such as the widely-used Roter Interaction Analysis System (RIAS).[6] RIAS offers a detailed framework for examining communication behaviors, particularly focusing on the socioemotional quality of interactions (i.e., "social signals",[8] which are verbal and nonverbal cues indicating attitudes, feelings, and relational stances). However, the process of manually coding these interactions within RIAS is labor-intensive. It is also prone to observer bias and presents significant challenges in scaling. These limitations underscore the potential for innovative approaches that can objectively, efficiently, and accurately assess communication dynamics in clinical settings.

Recent advancements in Artificial Intelligence (AI), notably in Automated Speech Recognition (ASR) and Natural Language Processing (NLP), offer promising avenues for addressing these challenges. The integration of ASR and NLP technologies has the potential to automate the identification and classification of social signals in clinical dialogs, such as interpersonal warmth or dominance, and can facilitate scalable assessment of patient-provider interactions,[9] with the potential to reveal hidden, implicit biases in clinical conversations. However, RIAS coding evaluates social signals by analyzing both non-verbal behaviors, such as eye contact and facial expressions, and elements of verbal communication, including the utterances that make up the dialogue, but not the full scope of the conversational content. To apply ASR and NLP for RIAS-based social signal predictions, such social cues would need to be accompanied by verbal indicators that can be detected using contemporary NLP methods. Analysis of the content of clinical conversations could provide a complement to nonverbally based analysis techniques for the detection of social signals.

This research builds upon previous research focused on the analysis of nonverbal clinical communication,[10] including the use of machine learning to characterize social signals in video recorded patient-provider interactions.[11] As a complement to nonverbal cues, which provide valuable context but do not consider communication content, transcribed speech offers a rich source of additional information about patient-provider dynamics. Nonverbal cues can indicate underlying emotions and attitudes. However, the content of verbal exchanges can convey expressions of concerns, beliefs, and needs that can inform our understanding of implicit biases, and how best to address them.[6] Therefore, the current research expands the scope of automated detection of social signals to the content of verbal communication, by applying ASR and NLP to classify social signals in clinical encounters.

**Methods**
*Dataset*
The audio data were collected from the Establishing Focus (EF) study.[12] Conducted between 2002 and 2006, the EF study was a randomized controlled trial of a brief intervention to increase physician skills at organizing and prioritizing encounter time with particular emphasis on using a patient-centered approach. The study generated audio recordings of 782 individual primary care patient visits from 782 different patients, with accompanying demographic data. In the original EF dataset, patient ethnicity categories included White, African American, Native American or Alaska Native, Asian, Native Hawaiian or other Pacific Islander, Hispanic, Other and Mixed. Among the participants represented in the dataset, 630 were White; 45 were Mixed; 36 were Asian; 33 were African American; 23 were Hispanic; 7 were categorized as Other; 5 were Native Hawaiian or other Pacific Islander and 4 were Native American or Alaska Native. For the purposes of this analysis, these groups were consolidated into two categories: "white" and "non-white," with the latter encompassing African American, Native American or Alaska Native, Asian, Native Hawaiian or other Pacific Islander, Hispanic, Other and Mixed classifications. This consolidation was performed to ensure a sufficient sample size for meaningful model training and statistical analysis across groups.

Table 1. RIAS social signals and coding definition.

| RIAS social signal | Signal definition |
| --- | --- |
| Dominance (assertiveness) | Confidently aggressive and self-assured, but can be domineering at times |
| Attentiveness (interest) | Showing interest in the concerns of others |
| Warmth (friendliness) | Showing kindness and caring |
| Engagement (responsiveness) | Being receptive to reacting appropriately to others |
| Empathy (sympathy) | Showing understanding and sensitivity toward the experiences of others |
| Respect | Being polite and showing grace to others |
| Interactivity (fully involved) | Keeping oneself involved in the interaction |
| Irritation (anger) | Showing feelings of displeasure or exasperation |
| Nervousness (anxiety) | Showing uneasiness or discomfort |
| Hurriedness* | Rushing through an interaction with speed and abrupt impatience |
| Sadness** (depression) | Showing unhappiness or sorrow |
| Emotional distress **(upset) | Showing emotional upset or suffering |

* Provider only signal, ** patient only signal

Within this dataset, 91 visits were manually coded using RIAS-based Global Affect Ratings, which we refer to as "social signals". These 91 coded visits, comprising 74 from white patients and 17 from non-white patients, were divided into 512 segments, each approximately 3 minutes in duration. A trained observer rated each 3-minute segment on a 1 ("low") to 6 ("high") scale for 12 social signals for both the patient and the provider as listed in Table 1. The remaining 691 uncoded visits, consisting of 556 from white patients and 135 from non-white patients, yielded 2,144 3-minute segments. Once we had validated the performance of our models using the coded visits, we applied the model with the best performance to these uncoded visits to assess differences in social signals across visits in which providers interacted with patients from different racial groups.

We observed extreme class imbalance where most rated signals fell near the mean rating for each signal. To address this issue and enhance the predictive capabilities of our models, we reformulated the task from a regression to a binary classification framework. To obtain binary labels, we viewed ratings as either "high" (above average) or "low" (below average) on their position relative to the average rating for the RIAS code concerned. Figure 1 shows the percentage of "high" labels (i.e. above average) for each social signal. Because of the homogeneity of the ratings, in some cases no examples, or only a few examples, fall below the average (for example, in provider-nervousness all the ratings are uniform with a value of one). Four signals have less than 2% of their labels in the "high" category: provider irritation, provider nervousness, provider hurriedness and patient empathy. This scarcity of data points led to their exclusion from the subsequent analysis, as the limited label diversity undermined the learning process. Furthermore, an additional 7 signals exhibited a highly skewed distribution, with a high label frequency of less than 10% in the training set, spanning from patient irritation (2%) to provider empathy (7.6%) as shown in Figure 1. These social signals included patient irritation, patient nervousness, patient respect, provider respect, patient sadness, patient distress, and provider empathy.

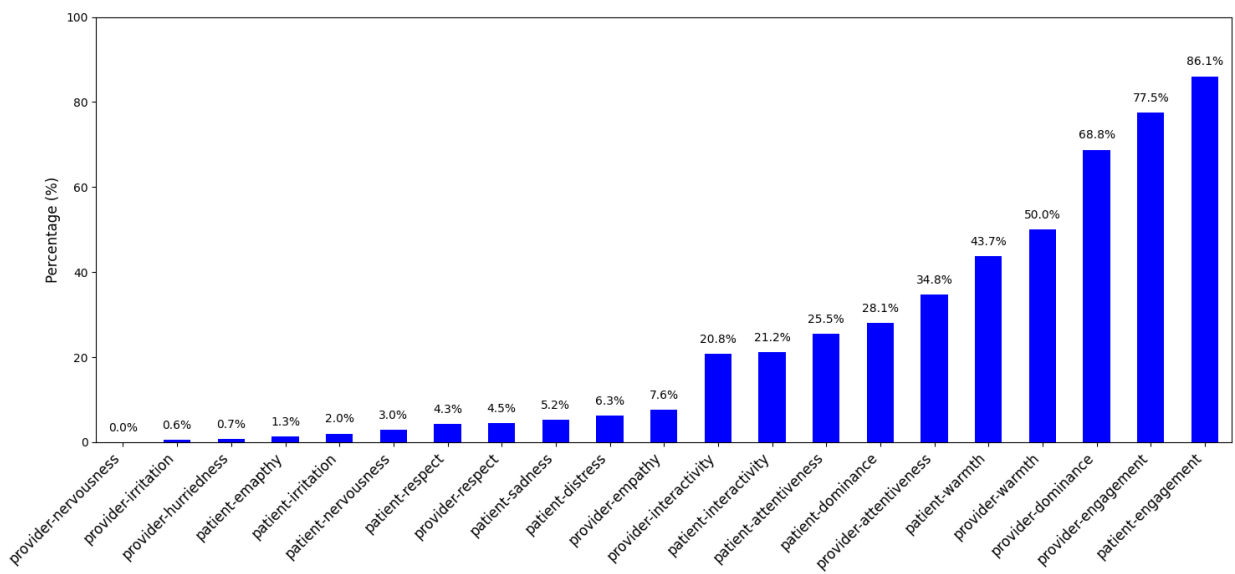

Figure 1. Percent of high social signal labels distribution across 3-minute segmentations

*Audio transcription and speech diarization*
The 91 RIAS coded visits were manually transcribed in the EF dataset and split into 3-minute segmentation based on the sentences' start time. For all visits in the EF study, OpenAI Whisper was used to transcribe the audio files.[13] Two model sizes were tested: large and large-v3. The Pyannote package[14, 15] was utilized for speaker diarization, which involves identifying and separating speakers within the audio, and for detecting the start time of each speaking turn. Further, the ASR transcribed text was segmented into 3-minute intervals, with segmentation beginning at the speech start time as determined by Pyannote's diarization process. To calculate the error rate of the ASR, visit-level automated transcriptions were compared with manual transcriptions after excluding time stamps, speaker labels, and punctuation from the analysis. Word Error Rate (WER) and Character Error Rate (CER) were calculated using the equation: $WER/CER = \frac{S + D + I}{N}$, where S is the number of substitutions, D is the number of deletions, I is the number of insertions, N is the number of words in the reference or the number of characters in the reference (where the reference is a manual transcription of the audio file).

*RIAS social signal prediction*
We observed extreme class imbalance where most scores for each signal fell close to the mean score, limiting the effectiveness of models that predict on a continuous scale. By converting to binary classification, we sought to emphasize the detection of distinctly high or low social signals, which are more likely to be indicative of meaningful interaction dynamics and potential biases[16]. Consequently, we cast social signal prediction as a binary categorization task, where the goal was to predict values that were above average. RIAS ratings were converted into 0 and 1 based on the aforementioned threshold determined by calculating the average across all segments within the training set for each signal respectively. Three Transformer-based NLP models were evaluated, on account of the strong performance of models based on this architecture on text categorization tasks.[17] For each of these models, a classifier was trained for each social signal (17 classifiers per model). As a baseline model, Sentence-BERT,[18] a variant of the BERT model that has been customized to enhance the similarity between embedding representations of sentences with related meaning, was used to embed the dialog. The resulting embeddings were provided to a multi-layer perceptron (MLP), which takes input of 768 features followed by a 256-neuron dense layer with ReLU activation and followed by a single-neuron output layer using sigmoid activation function to predict binarized RIAS scores. In addition, we evaluated two models using end-to-end fine-tuning (instead of training an MLP for classification only): RoBERTa[19] and BERT.[20] A class-weighted binary cross entropy loss function was used to mitigate the effects of differences in the number of high examples of each code in the training set. We assign a weight to each class that is inversely proportional to its frequency in the dataset. This weight is computed as the total number of transcripts divided by the number of transcripts that belong to the class we're focusing on. We performed 5-fold cross-validation and averaged the evaluation metrics of each model across the five folds. Accuracy, precision, recall, weighted F-1 scores, Area Under the Receiver Operating Characteristic Curve) AUROC and Area Under the Precision-Recall Curve (AUPRC) were calculated.

*Fairness assessment*
In order to draw conclusions about differences in social signals across groups, it is necessary to establish that our models do not confound our measurements by differing in accuracy across these groups. We assessed the fairness of our ASR pipeline by conducting a T-test to compare WER across all manually transcribed visits between white and non-white patient groups. Furthermore, to assess potential confounders, we conducted a chi-square test to examine the gender distribution within the racial groups in both coded and uncoded datasets, ensuring gender did not disproportionately influence our findings.

Demographic parity differences between white and non-white groups were assessed for each signal prediction. This metric provides an intuitive measure of whether different groups receive positive outcomes at similar rates. We selected this demographic parity because it is applicable to imbalanced datasets, like ours. To examine potential disparities and assess the fairness of our model's predictions, we applied a bootstrap method to the 3-minute segmented interactions within the test set, which was split from the coded dataset. This involved resampling the data 1000 times to create a distribution of demographic parity differences. For each resample, we calculated the differences in prediction outcomes between racial groups across these segments. A 95% confidence interval was then established from this distribution to determine if there was a statistically significant disparity that could indicate model bias. This method allows us to determine whether our model's predictions disproportionately favor one group over another. This provides insight into the model's fairness in predicting outcomes for different racial groups.

*Social signal disparity detection*
To obtain a summary representation of each signal for each visit, we averaged the predicted scores across all segments derived from it. Then, a Mann–Whitney U test was performed on the predicted binarized RIAS signals for the larger uncoded dataset, to detect statistically significant differences in RIAS social signals. This test was chosen because it does not assume a normal distribution of the scores and is suitable for ordinal data with unequal sample sizes between groups.

**Results**
*ASR error rate*
The 'large' model configuration resulted in a WER of 55.4% and a CER of 42.2%, indicating the proportion of word and character errors made by the model compared to a manual transcription. In contrast, the 'large-v3' configuration displayed a better performance, reducing the WER to 43.3% and the CER to 31.8%. It should be noted that these error rates were derived by comparing the ASR output against manually transcribed text that had been de-identified. This de-identification process may omit certain elements of the audio content, thereby potentially inflating the calculated error rates. Thus, the reported error rates might overestimate the actual performance of the ASR system, as the manual transcriptions do not capture the entirety of the spoken content in the recordings.

*NLP model comparison on manual transcribed text*

In our analysis, three distinct NLP models were utilized: SBERT+MLP, RoBERTa, and BERT, with fine-tuned variants of each model tasked with predicting one of the 17 social signals. Figure 2 illustrates a side-by-side comparison of these models across AUPRC The aggregated performance metrics with standard deviation for the Sentence-BERT with a Multi-Layer Perceptron (SBERT + MLP), BERT, and RoBERTa models are summarized in Table 2.

Table 2. NLP model performance (five-fold cross-validation)

|  | SBERT + MLP (mean ± sd) | RoBERTa (mean ± sd) | BERT (mean ± sd) |
|---|---|---|---|
| **Accuracy** | 0.762 ± 0.131 | 0.861 ± 0.076 | ***0.906 ± 0.059*** |
| **Weighted F1** | 0.759 ± 0.138 | 0.854 ± 0.070 | ***0.899 ± 0.055*** |
| **AUROC** | 0.600 ± 0.060 | 0.680 ± 0.143 | ***0.737 ± 0.155*** |
| **AUPRC** | 0.387 ± 0.264 | 0.436 ± 0.341 | ***0.543 ± 0.339*** |

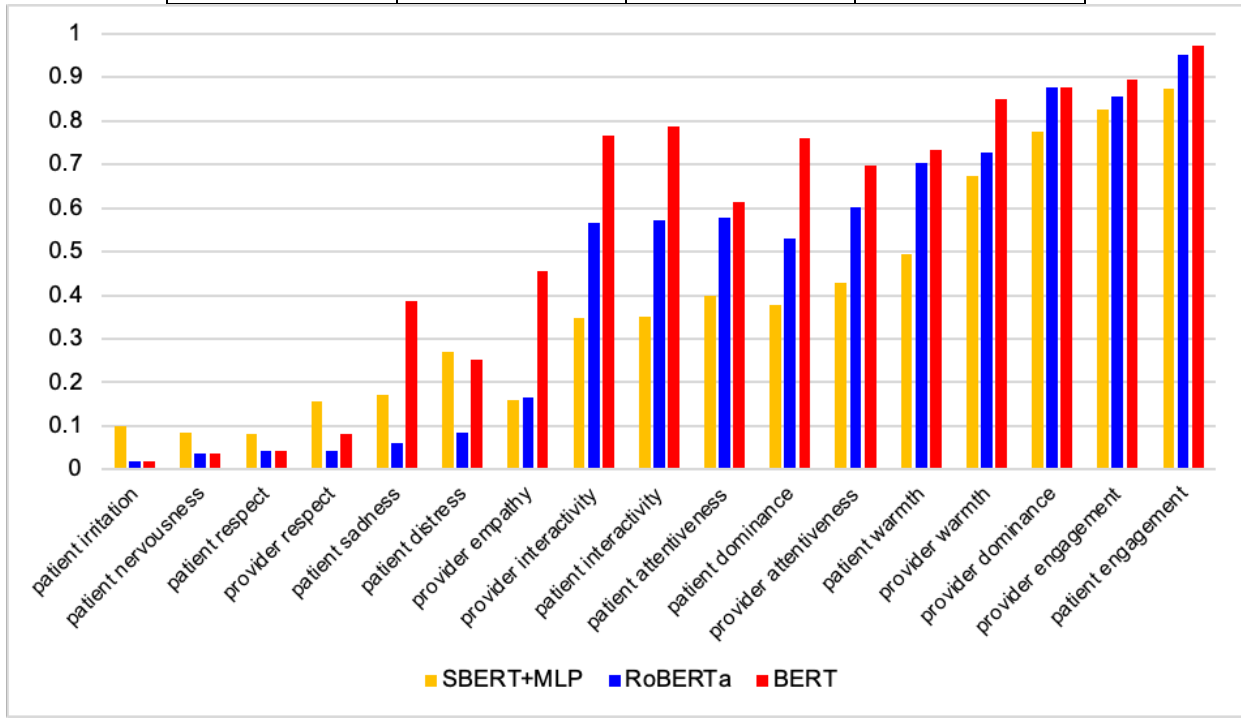

Figure 2. Comparison of AUPRC among NLP models. Codes ordered by ascending frequency of positive examples.

The BERT model demonstrated superior performance across all evaluated metrics when compared to the other models. However, for four social signals—patient irritation, patient nervousness, patient respect, and provider respect (leftmost bars in Figure 2) —BERT's AUPRC was observed to be less than 0.1. The diminished performance of our models can be attributed to the highly imbalanced distribution of labels in each signal. All these four signals have less than 10% of high labels. This imbalance meant that the minority class was underrepresented, hindering the effective training of the models.

*Comparison of model performance on ASR-transcribed text*

We proceeded to assess the best-performing model, BERT, on text transcribed automatically by ASR. The comparative analysis, summarized in Table 3, reveals that the model's predictive performance on ASR-transcribed text closely mirrors that on manual transcriptions. Surprisingly, the AUROC and AUPRC values showed a significant increase when using ASR-generated text. The four signals with AUPRC below 0.1 using manual transcripts had up to threefold increases in AUPRC with ASR, as shown in Figure 3. Interestingly, AUROC and AUPRC improved most when using the 'large' configuration with a higher word and character error rate — average AUROC and AUPRC reached 0.798 and 0.668 respectively. This suggests that the model may benefit from the variations inherent in ASR transcriptions, as was discovered in previous studies of participants experiencing hallucinations and dementia patients.[21] These results suggest that our BERT classifier is robust to ASR errors, and may even benefit from them.

Table 3. Performance comparison with manual and ASR-derived transcriptions

|  | Manual (mean ± sd) | ASR (mean ± sd) |
| --- | --- | --- |
| **Accuracy** | ***0.906 ± 0.059*** | 0.901 ± 0.065 |
| **Weighted F1** | ***0.899 ± 0.055*** | 0.898 ± 0.060 |
| **AUROC** | 0.737 ± 0.155 | ***0.763 ± 0.094*** |
| **AUPRC** | 0.543 ± 0.339 | ***0.603 ± 0.239*** |

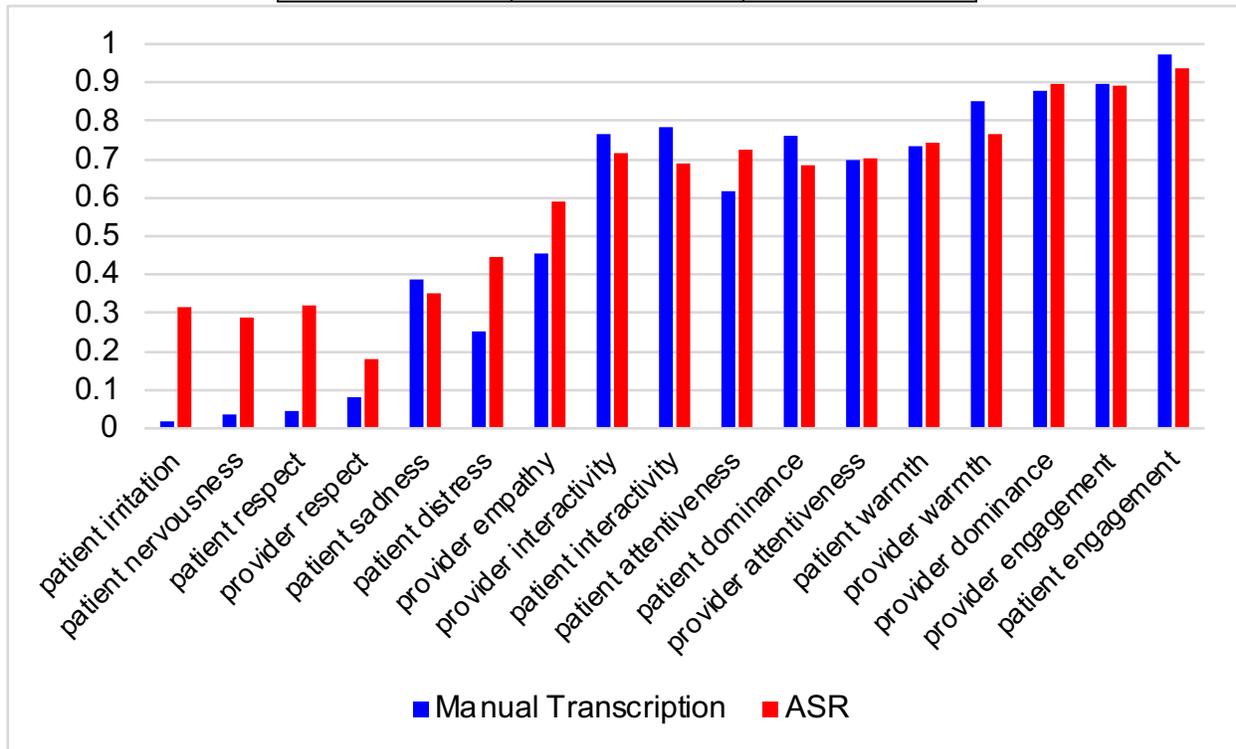

Figure 3. Comparison of AUPRC for manual and ASR-derived transcriptions

*Evaluation of pipeline fairness*
In examining the fairness of the automated pipeline, we first evaluated the ASR error rates to ascertain whether the transcription process introduced any biases based on patient race. 75 white patients had WER (M = 0.511, SD = 0.217) compared to the 16 non-white patients (M = 0.443, SD = 0.103), demonstrating no significant effect for race, t (89) = 1.23, p=0.222. Hence, this indicates balanced ASR performance across white and non-white groups.

To further ensure that findings were not confounded by gender distribution differences across racial groups, we conducted a chi-square test. We found no significant difference in gender distribution between white and non-white patients in both the coded dataset (chi-square statistic $\chi^2$ (1, N = 91) = 0.207, p = 0.65 > 0 .05) and the uncoded dataset (chi-square statistic $\chi^2$ (1, N = 691) = 0.214, p = 0.64 > 0 .05). That gender distribution across racial groups did not significantly differ in our sample, indicates that gender does not act as a confounder in our subsequent analysis of group differences.

All 17 models used for signal classifications were statistically tested to examine the demographic parity differences between the two groups. A value close to 0 indicates little to no disparity, implying fairness in the predictions made by the models. For each signal, we calculated the mean demographic parity difference and constructed a 95% confidence interval for these differences, providing a statistical significance measure for the disparities observed, which is detailed in Table 4. The test found that 0 was within the 95% interval for all signals, suggesting that there is no statistically significant difference between two groups (Table 3). Together, these results suggest that our automated pipeline performs predictions that are fair and unbiased across white and non-white racial groups. Therefore, any group differences in RIAS codes identified cannot be attributed to differences in pipeline performance across these groups.

Table 4. Demographic parity differences and confidence interval between white and non-white patient

| Social Signal | Mean Demographic Parity Difference* | 95% Confidence Interval |
|---|---|---|
| patient irritation | -0.001 | (-0.097, 0.062) |
| patient nervousness | -0.002 | (-0.097, 0.062) |
| patient respect | 0.005 | (-0.089, 0.078) |
| provider respect | -0.040 | (-0.120, 0.000) |
| patient sadness | -0.044 | (-0.212, 0.100) |
| patient distress | -0.003 | (-0.148, 0.124) |
| provider empathy | 0.018 | (-0.139, 0.145) |
| provider interactivity | 0.035 | (-0.160, 0.215) |
| patient interactivity | 0.020 | (-0.175, 0.198) |
| patient attentiveness | 0.063 | (-0.126, 0.229) |
| patient dominance | -0.115 | (-0.313, 0.079) |
| provider attentiveness | -0.131 | (-0.345, 0.086) |
| patient warmth | -0.002 | (-0.207, 0.215) |
| provider warmth | 0.095 | (-0.112, 0.285) |
| provider dominance | -0.002 | (-0.175, 0.193) |
| provider engagement | -0.109 | (-0.238, 0.045) |
| patient engagement | -0.107 | (-0.247, 0.055) |

*Positive demographic parity difference means the model predicts positive outcomes at a higher rate for white patients than for non-white patients; negative difference means the model predicts positive outcomes at a lower rate for white patients than for non-white patients.

*Analysis of differences in social signals in the EF dataset*

Differences in social signal rating between white and non-white groups were detected using the Mann-Whitney U tests from the entire uncoded dataset, z-score and p-value were reported for each signal in Table 5. Statistically significant differences were detected for seven RIAS social signals. Compared with non-white patients, white patients were more likely to receive higher levels of provider warmth ($z = 2.31$, $p = 0.009$), suggesting a friendlier attitude towards white patients than non-white patients. White patients also experienced greater expression of provider engagement ($z = 1.86$, $p = 0.020$) and provider attentiveness ($z = 1.64$, $p = 0.039$), which might indicate more receptive and attentive care. The average word count ratio between provider and patient in the coded dataset also support this finding: providers' word count is nearly three times higher with white patients than with non-white patients. Furthermore, provider dominance was notably higher towards white patients ($z = 2.96$, $p = 0.001$), potentially implying a more authoritative interaction style, which is consistent with previous findings from our group.[22] In contrast, previous works report higher levels of dominance expressed by providers when interacting with Black patients.[7, 23] One explanation for this discrepancy may relate to the observation that white patients displayed higher levels of nervousness ($z = 0.49$, $p = 0.047$), which was paralleled by greater patient engagement ($z = 1.43$, $p = 0.049$). This could result in a heightened responsiveness of providers to white patients' emotional states resulting in a higher provider dominance. Interestingly, our analysis suggests that white patients experienced lower levels of provider interactivity ($z = -1.88$, $p = 0.011$). A potential explanation for this is providers were attentive listeners, taking longer pauses to process patient's input before responding, which could be interpreted as more patient-centered communication.

Table 5. Social signals statistical differences between white and non-white patient

| Social Signal | z-score* | p-value** |
|---|---|---|
| patient irritation | 0.10 | 0.197 |
| patient nervousness | *0.49* | *0.047* |
| patient respect | 0.05 | 0.452 |
| provider respect | -0.24 | 0.098 |
| patient sadness | 0.16 | 0.220 |
| patient distress | 0.36 | 0.161 |
| provider empathy | -0.14 | 0.396 |
| provider interactivity | *-1.88* | *0.011* |
| patient interactivity | -0.99 | 0.106 |
| patient attentiveness | 0.92 | 0.102 |
| patient dominance | -0.51 | 0.246 |
| provider attentiveness | *1.64* | *0.039* |
| patient warmth | 1.34 | 0.068 |
| provider warmth | *2.31* | *0.009* |
| provider dominance | *2.96* | *0.001* |
| provider engagement | *1.86* | *0.020* |
| patient engagement | *1.43* | *0.049* |

\* A positive z-score indicates a higher average rating for white patients compared to non-white patients; a negative z-score indicates a lower average rating for white patients compared to non-white patients.
\*\*P-values highlighted in bold are less than 0.05, denoting statistical significance.

**Discussion**

Improving patient-provider interactions is vital for patient-centered, equitable healthcare that improves patient satisfaction and outcomes. RIAS is a useful tool to assess disparities in social signals within the socioemotional context of clinical conversations and has been shown to be associated with implicit race bias.[7] However, traditional applications of RIAS requires manual coding, which cannot be easily scaled. In this study, we built and evaluated the performance of an automated pipeline to make use of verbal information in recorded primary care visits to predict social signals, and identified differences in these social signals that may reflect bias in patient-provider interactions. This suggests that the non-verbal cues used to assign RIAS codes are accompanied by linguistic indicators of social signals that can be used to identify them.

We successfully developed an automated pipeline for transcribing audio recordings of clinical visits using ASR and predicting social signal scores using NLP, achieving an average accuracy of 90.1% and an AUC of 76.3% for 17 RIAS signals. This demonstrates the robustness of the automated system in capturing nuances in patient-provider interactions. Notably, we found no significant differences in the error rate of ASR or differences in demographic parity between white and non-white patient visits, demonstrating the fairness of the automated pipeline.

Seven signals with statistically significant differences in communication between visits with white and non-white patient were identified. These findings corroborate existing literature that suggests health care providers often harbor biases, consciously (i.e., explicit bias) or unconsciously (i.e., implicit bias), resulting in differential treatment of patients based on race.[4, 5] The fact that providers exhibited more positive attitudes towards white patients than non-white patients aligns with research showing evidence that health care providers hold implicit race bias, similar to others in society.[5, 24] Conversely, the higher incidence of anxiety-related signals in white patients suggests a complex interplay of factors that warrants further investigation.

Findings indicating that providers tend to express more warmth, engagement, and interest when interacting with white patients than non-white patients are particularly concerning. These differences in provider communication behavior may represent hidden biases that could potentially contribute to the well-documented disparities in healthcare outcomes. Cooper et al.[7] found that provider verbal dominance increased with both black patients (9%) and white patients (11%) with each 0.5 increase in provider implicit race bias score. In the current work, the greater provider dominance observed towards white patients may reflect a compensatory mechanism for the heightened nervousness these patients expressed, suggesting that providers may instinctively exert control in situations perceived as stressful.

These findings highlight the complex nature of detecting differences in clinical communication that may represent bias and underscore the promise of using such automated analyses to identify and understand these biases, which are often subtle and unconscious, yet can significantly impact the quality of patient-provider interaction and healthcare outcomes. Our automated pipeline presents an opportunity to uncover and address potential bias in clinical interactions systematically. This integration highlights the role of advanced computational methods in enhancing the patient-centered healthcare environments, thereby improving patient care through the identification and mitigation of bias. By highlighting specific social signals that may be associated with implicit bias, targeted interventions can be developed to educate providers in recognizing and mitigating their biases.

**Limitations and future work**
The data set, derived from the EF study conducted between 2002 and 2006, may not fully represent current primary care clinical communication practices. The dates of the clinical interactions may affect the generalizability of the model to contemporary data, on account of the temporal gap. Additionally, the occurrence of some outcomes, such as patient irritation and patient nervousness, in the training data may be too low to evaluate effectively in the remaining dataset. Also, the binary classification of social signals as below or above average may oversimplify the complex spectrum of social interactions. Furthermore, the imbalance in the number of white and non-white patient visits could influence the predictive modeling and subsequent bias analysis. Our model focused on the verbal behaviors in the recording. Non-verbal behaviors, such as eye contact or pauses, were not included as features. The unexpected finding that higher error rates in ASR might improve AUROC and AUPRC poses additional questions about the relationship between transcription accuracy and predictive modeling, to be evaluated in future work. The demographic data for providers were not included in the analysis. As such we were unable to assess patient-provider concordance in identity, which has been associated with better communication.[25] Future work should account for provider race, and also explore racially concordant versus discordant visits. In addition, we plan to evaluate our pipeline using a larger data set, and examine more granular racial categories to describe interactions with patients from historically marginalized groups and other demographic categories (gender, ethnicity, age, etc.), combine verbal and nonverbal information for a more holistic view of clinical interactions, and use explainable AI techniques to identify extract keywords or phrases in the conversation that contribute to the model predictions.

**Conclusion**
We developed an automated pipeline that employs ASR and NLP to classify RIAS social signals. By applying the validated classifier to a larger, uncoded dataset, our work sets the stage for comparing social signals in clinician interactions across diverse patient groups at scale, a significant step forward in understanding and addressing biases in healthcare communication. This work enables a deep analysis of verbal patient-provider communication and facilitates the potential identification of implicit bias in communications. By advancing the assessment and understanding of biased healthcare communication, this study contributes to the broader goals of improving patient-provider communication quality, promoting healthcare equity, and ensuring that all patients receive respectful and empathetic care during clinical visits.

**Acknowledgments**
This study was supported in part by NIH R01LM013301 and R01LM014056-02S1. Additionally, we wish to acknowledge the invaluable efforts of the students who coded the EF dataset: Kimberly Sladek, Alexandra Andreiu, Kelly Tobar, Veen Doski, and Anuujin Tsedenbal. Their dedication and hard work have significantly contributed to the richness and depth of our analysis. We also wish to thank Dr. Debra Roter for her guidance on manual coding of recorded visits with RIAS Global Affect Ratings.

References

1. Patak L, Wilson-Stronks A, Costello J, et al. Improving patient-provider communication: a call to action. The Journal of nursing administration. 2009;39(9):372.
2. Kwame A, Petrucka PM. A literature-based study of patient-centered care and communication in nurse-patient interactions: barriers, facilitators, and the way forward. BMC nursing. 2021;20(1):1-10.
3. Kang J, Bennett M, Carbado D, et al. Implicit bias in the courtroom. UCLa L rev. 2011;59:1124.
4. FitzGerald C, Hurst S. Implicit bias in healthcare professionals: a systematic review. BMC medical ethics. 2017;18(1):1-18.
5. Hall WJ, Chapman MV, Lee KM, et al. Implicit racial/ethnic bias among health care professionals and its influence on health care outcomes: a systematic review. American journal of public health. 2015;105(12):e60-e76.
6. Hagiwara N, Lafata JE, Mezuk B, et al. Detecting implicit racial bias in provider communication behaviors to reduce disparities in healthcare: challenges, solutions, and future directions for provider communication training. Patient education and counseling. 2019;102(9):1738-43.

7.	Cooper LA, Roter DL, Carson KA, et al. The associations of clinicians' implicit attitudes about race with medical visit communication and patient ratings of interpersonal care. American journal of public health. 2012;102(5):979-87.
8.	Burgoon JK, Magnenat-Thalmann N, Pantic M, Vinciarelli A. Social signal processing: Cambridge University Press; 2017.
9.	van Buchem MM, Boosman H, Bauer MP, et al. The digital scribe in clinical practice: a scoping review and research agenda. NPJ digital medicine. 2021;4(1):57.
10.	Hartzler A, Patel R, Czerwinski M, et al. Real-time feedback on nonverbal clinical communication. Methods of information in medicine. 2014;53(05):389-405.
11.	Bedmutha MSaT, Anuujin and Tobar, Kelly and Borsotto, Sarah and Sladek, Kimberly R and Singh, Deepansha and Casanova-Perez, Reggie and Bascom, Emily and Wood, Brian and Sabin, Janice and Pratt, Wanda and Hartzler, Andrea and Weibel, Nadir. ConverSense: An Automated Approach to Assess Patient-Provider Interactions using Social Signals. Proceedings of the 2024 CHI Conference on Human Factors in Computing Systems. 2024.
12.	Brock DM, Mauksch LB, Witteborn S, et al. Effectiveness of intensive physician training in upfront agenda setting. Journal of General Internal Medicine. 2011;26:1317-23.
13.	Radford A, Kim JW, Xu T, et al., editors. Robust speech recognition via large-scale weak supervision. International Conference on Machine Learning; 2023: PMLR.
14.	Plaquet A, Bredin H. Powerset multi-class cross entropy loss for neural speaker diarization. arXiv preprint arXiv:231013025. 2023.
15.	Bredin H, editor pyannote. audio 2.1 speaker diarization pipeline: principle, benchmark, and recipe. 24th INTERSPEECH Conference (INTERSPEECH 2023); 2023: ISCA.
16.	Skinner AL, Meltzoff AN, Olson KR. "Catching" social bias: Exposure to biased nonverbal signals creates social biases in preschool children. Psychological Science. 2017;28(2):216-24.
17.	Minaee S, Kalchbrenner N, Cambria E, et al. Deep learning--based text classification: a comprehensive review. ACM computing surveys (CSUR). 2021;54(3):1-40.
18.	Reimers N, Gurevych I. Sentence-bert: Sentence embeddings using siamese bert-networks. arXiv preprint arXiv:190810084. 2019.
19.	Liu Y, Ott M, Goyal N, et al. Roberta: A robustly optimized bert pretraining approach. arXiv preprint arXiv:190711692. 2019.
20.	Devlin J, Chang M-W, Lee K, Toutanova K. Bert: Pre-training of deep bidirectional transformers for language understanding. arXiv preprint arXiv:181004805. 2018.
21.	Li C, Xu W, Cohen T, Pakhomov S. Useful blunders: Can automated speech recognition errors improve downstream dementia classification? Journal of Biomedical Informatics. 2024;150:104598.
22.	Manas Satish Bedmutha AB, , Sabrina Mangal, Emily Bascom, Wanda Pratt, Brian Wood, Janice Sabin, MSW2, Nadir Weibel, Andrea Hartzler. Towards inferring implicit bias in clinical interactions using social signals. American Medical Informatics Association 2023 AI Showcase. 2023.
23.	Johnson RL, Roter D, Powe NR, Cooper LA. Patient race/ethnicity and quality of patient–physician communication during medical visits. American journal of public health. 2004;94(12):2084-90.
24.	Nosek BA, Smyth FL, Hansen JJ, et al. Pervasiveness and correlates of implicit attitudes and stereotypes. European review of social psychology. 2007;18(1):36-88.
25.	Shen MJ, Peterson EB, Costas-Muñiz R, et al. The effects of race and racial concordance on patient-physician communication: a systematic review of the literature. Journal of racial and ethnic health disparities. 2018;5:117-40.